\documentclass[10pt,twocolumn,showpacs,preprintnumbers,amsmath,amssymb,aps,prl,superscriptaddress,longbibliography]{revtex4-2}
\usepackage{appendix}
\usepackage{bbm}
\usepackage{mathrsfs} 
\usepackage{graphicx}
\usepackage{dcolumn}
\usepackage{bm}
\usepackage{braket}
\usepackage{amsmath}
\usepackage{amsfonts}
\usepackage{CJKutf8}
\usepackage[dvipsnames]{xcolor}
\usepackage[colorlinks=true,linkcolor=Blue,urlcolor=BlueViolet,citecolor=BlueViolet]{hyperref}
\usepackage{natbib}
\usepackage{multirow}

\begin{document}

\title{Intrinsic Quantum Mpemba Effect in Markovian Systems and Quantum Circuits}
\author{Dongheng Qian}
\affiliation{State Key Laboratory of Surface Physics and Department of Physics, Fudan University, Shanghai 200433, China}
\affiliation{Shanghai Research Center for Quantum Sciences, Shanghai 201315, China}
\author{Huan Wang}
\affiliation{State Key Laboratory of Surface Physics and Department of Physics, Fudan University, Shanghai 200433, China}
\affiliation{Shanghai Research Center for Quantum Sciences, Shanghai 201315, China}
\author{Jing Wang}
\thanks{wjingphys@fudan.edu.cn}
\affiliation{State Key Laboratory of Surface Physics and Department of Physics, Fudan University, Shanghai 200433, China}
\affiliation{Shanghai Research Center for Quantum Sciences, Shanghai 201315, China}
\affiliation{Institute for Nanoelectronic Devices and Quantum Computing, Fudan University, Shanghai 200433, China}
\affiliation{Hefei National Laboratory, Hefei 230088, China}

\begin{abstract}

The quantum Mpemba effect (QME) describes the counterintuitive phenomenon in which a system farther from equilibrium reaches steady state faster than one closer to equilibrium. However, ambiguity in defining a suitable distance measure between quantum states has led to varied interpretations across different contexts. Here we propose the intrinsic quantum Mpemba effect (IQME), defined using the trajectory length traced by the quantum state as a more appropriate measure of distance—distinct from previous trajectory-independent metrics. By treating quantum states as points in a Riemannian space defined by statistical distance, the trajectory length emerges as a more natural and accurate characterization of the counterintuitive dynamics, drawing an analogy to the classical Brachistochrone problem.  We demonstrate the existence of IQME in Markovian systems and extend its definition to quantum circuits, thereby establishing a unified framework applicable to both open and closed systems. Notably, we observe an IQME in a $U(1)$-symmetric circuit, offering new insights into the rates of quantum thermalization for different initial states. This work deepens our understanding of quantum state evolution and lays the foundation for accurately capturing novel quantum dynamical behaviour.

\end{abstract}


\maketitle

The classical Mpemba effect, first observed over half a century ago, describes the surprising phenomenon where warmer water can freeze more quickly than colder water~\cite{mpemba1969, jeng2006}. This effect has recently garnered renewed attention~\cite{lu2017, klich2019a, kumar2020, gal2020} and has been extended to the quantum domain~\cite{warring2024}, where a quantum Mpemba effect (QME) has been both theoretically predicted and experimentally verified~\cite{shapira2024, zhang2024d, joshi2024}. This generalization applies to both open quantum systems, especially those subject to Markovian dissipation, and closed systems, in which parts of the system itself act as a thermal bath. Notably, the QME has also been observed within a quantum circuit model~\cite{liu2024c, liu2024d, turkeshi2024}, a minimal framework for approximating closed-system evolution.

The counterintuitive essence of the QME, applicable to both open and closed quantum systems, lies in the paradoxical behavior where a quantum state initially ``farther'' from equilibrium reaches the final state more rapidly. Characterizing this phenomenon requires a predefined measure of distance between quantum states, and various distance measures have been utilized in previous studies. For Markovian open systems, measures such as trace distance~\cite{bao2022a, furtado2024, ivander2023, longhi2024a}, Hilbert-Schmidt distance~\cite{carollo2021}, Kullback-Leibler divergence~\cite{lapolla2020, longhi2024, manikandan2021, moroder2024}, Euclidean distance in parameter space~\cite{nava2024}, quantum mutual information~\cite{wang2024c}, inverse participation ratio~\cite{dong2024}, effective temperature~\cite{wang2024d}, and differences between matrix elements~\cite{chatterjee2023,chatterjee2024} have been employed. For closed systems, entanglement asymmetry has become the most widely used measure~\cite{ares2023,ares2023a, ares2023b,ares2024,caceffo2024,chalas2024,ferro2024,rylands2024,rylands2024a,yamashika2024,yamashika2024a}. This diversity in definitions raises the important question of which distance measure is most appropriate and whether different measures yield consistent results. A recent study highlighted this ambiguity and proposed the thermomajorization Mpemba effect to provide a unified definition~\cite{vu2024}. Nonetheless, a critical limitation shared by all these measures is their trajectory independence; that is, the distance measure does not depend on the specific path taken by the quantum state during its evolution. Intuitively, even if a state is geodesically closer to the final state, it may still take more time to reach it if the trajectory is longer.

\begin{figure}[t]
\begin{center}
\includegraphics[width=3.35in, clip=true]{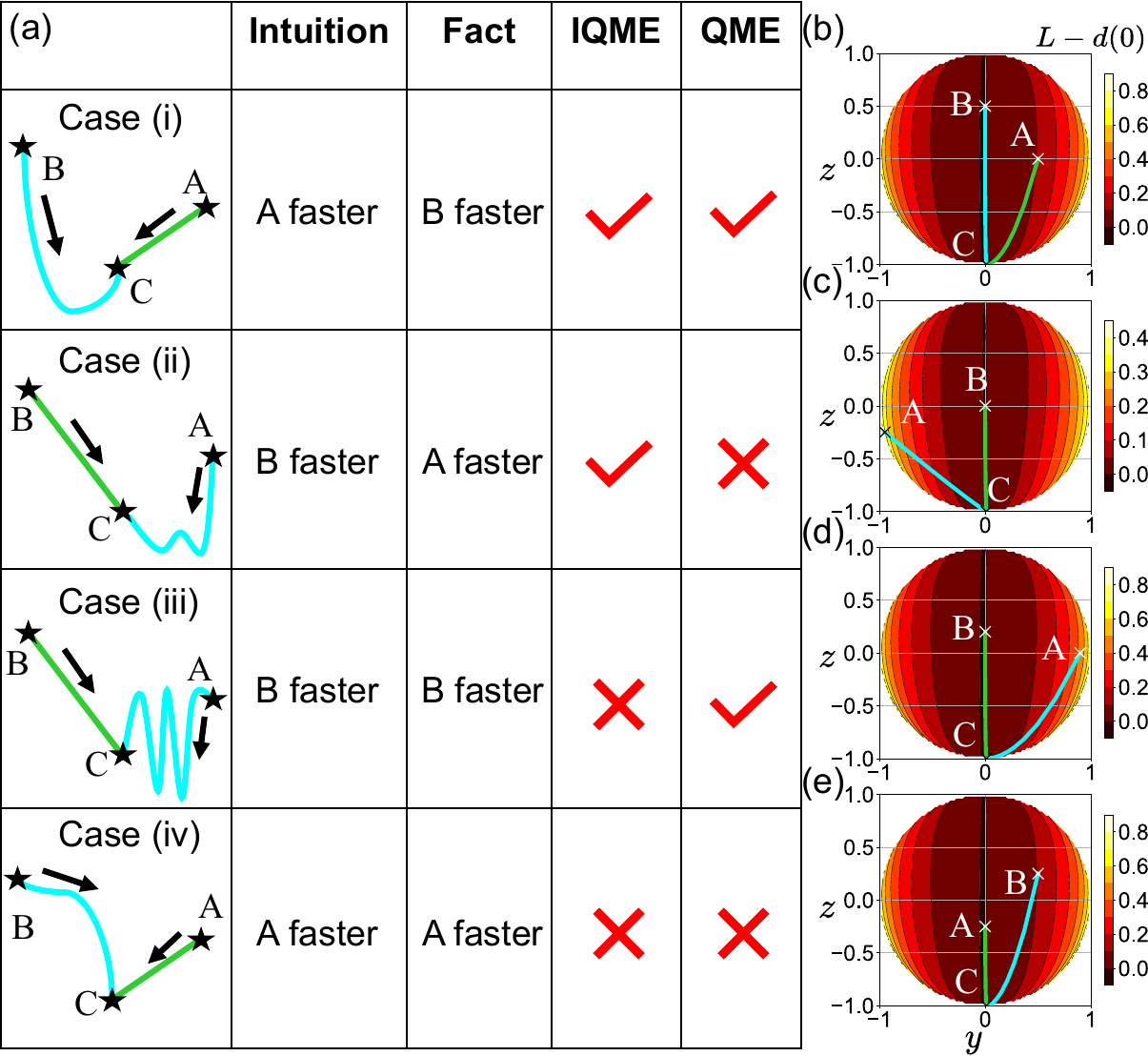}
\end{center}
\caption{Comparison for intuition, fact, IQME and QME. (a) The leftmost column presents four cases, which can be analogized to particles moving in a classical gravitational field, where one might intuitively expect that a greater trajectory distance leads to a longer travel time. In each case, the longer trajectory is shown in cyan, while $A$ is always geodesically closer to the steady state. However, as in the counterintuitive Brachistochrone problem, these intuitions can be misleading. The occurrence of IQME corresponds precisely to scenarios where intuition fails. (b-e) Quantum state evolution under Eq.~(\ref{eq2}) with $\alpha=100$; $\gamma’=0.94$ in (b,d,e) and $\gamma’=0.52$ in (c). (b) $L_A = 0.890 < L_B = 1.046$, while $d_A(0) = 0.782 < d_B(0) = 1.046$, corresponding to case (i). (c) $L_A = 1.019 > L_B = 0.781$, while $d_A(0) = 0.663 < d_B(0) = 0.781$, corresponding to case (ii). (d) $L_A = 1.214 > L_B = 0.885$, while $d_A(0) = 0.780 < d_B(0) = 0.885$, corresponding to case (iii). (e) $L_A = 0.658 < L_B = 1.013$, while $d_A(0) = 0.658 < d_B(0) = 0.908$, corresponding to case (iv).}
\label{fig1}
\end{figure}

In this Letter, we introduce the intrinsic quantum Mpemba effect (IQME), defined by using the trajectory length traced by the quantum state during its evolution as the measure of distance. By conceptualizing quantum states as points on a Riemannian manifold, where the metric is defined by statistical distinguishability, we can compute the trajectory distance straightforwardly according to this metric. The IQME describes a scenario in which a quantum state $\rho_A$, despite having a longer trajectory distance to a target state $\rho_C$ than another state $\rho_B$, reaches $\rho_C$ in a shorter time. This mirrors the classical Brachistochrone problem, where a longer path can lead to a shorter travel time. In contrast, the original QME uses geodesic distance as its metric. While the IQME and QME can occur independently, the IQME offers a more accurate representation of counterintuitive dynamics. As shown in Fig.~\ref{fig1}(a), although point $A$ has a shorter geodesic distance to the target than point $B$ in all cases, intuition suggests that we should compare the actual trajectory length, 
with the longer path depicted by the cyan line for clarity. Thus, counterintuitive behavior aligns naturally with the IQME, while the original QME may be trivially expected due to the longer trajectory taken by the geodesically closer state in some cases. Consequently, the IQME offers a more precise and appropriate characterization of the peculiar dynamics underlying the Mpemba effect. To the best of our knowledge, while recent studies have introduced a similar information geometry perspective on the QME~\cite{bettmann2024,bravetti2024,srivastav2024}, a detailed definition and demonstration of the IQME has yet to be proposed.

We first confirm the presence of the IQME in a Markovian system, using a model in which an inverse QME has been experimentally verified. We then generalize the definition to closed system, providing a unified framework for the IQME. To illustrate this, we examine a $U(1)$-symmetric quantum circuit, where the IQME appears for tilted N\'eel initial states but is absent for tilted ferromagnetic states—a result that contrasts with the original QME, defined via entanglement asymmetry~\cite{liu2024c,turkeshi2024}. 
This observation not only explains the QME behavior in tilted ferromagnetic initial states but also suggests that faster symmetry restoration does not necessarily correlate with a higher evolution speed, offering new insights into the intricate relationship between symmetry and thermalization in quantum systems.

\emph{IQME definition.---}The representation of classical probability distributions as points on a manifold is well-established in the field of information geometry~\cite{nielsen2020}, a framework that has since been extended to quantum states~\cite{wootters1981}. A fundamental requirement for a valid metric on quantum states is monotonicity, which ensures that the metric remains contractive under any completely positive trace-preserving map $\Phi$. Specifically, the distance $ds^{2} \equiv \mathcal{D}(\rho, \rho + d\rho)$ between $\rho$ and $\rho + d\rho$ must satisfy  $\mathcal{D}(\Phi(\rho), \Phi(\rho + d\rho)) \leq \mathcal{D}(\rho, \rho + d\rho)$. This metric is commonly referred to as quantum Fisher information in the study of quantum metrology~\cite{petz2011a,scandi2024,marvian2022,giovannetti2006}. Petz~\cite{petz1996} has shown that any metric satisfying this requirement can be expressed in the following form:
\begin{equation}
\mathcal{D}(\rho, \rho+d\rho) = \text{Tr}[d\rho \mathcal{J}_f^{-1}|_{\rho}(d\rho)],
\end{equation}
where $\mathcal{J}_f |_\rho = \mathbb{R}_{\rho}f(\mathbb{L}_{\rho}\mathbb{R}_{\rho}^{-1})$, $\mathbb{R}_{\rho}[A] = A\rho$ and $\mathbb{L}_{\rho}[A] = \rho A$. The matrix function $f$ must be matrix monotone and satisfy $f(x) = xf(x^{-1})$ and $f(1)=1$ to ensure that $\mathcal{D}(\rho, \rho + d\rho)$ is symmetric and positive. A particularly popular choice is $f(x) = (x+1)/2$, known as the Bures metric~\cite{bures1969, braunstein1994}. For this metric, the geodesic distance between two quantum states $\rho$ and $\sigma$ is given by the analytical expression $\mathcal{D}^{\text{geo}}(\rho, \sigma) = 2\text{arccos}(F(\rho,\sigma))$, where $F(\rho, \sigma)$ is the Uhlmann fidelity, defined as $F(\rho, \sigma) = \text{Tr}(\sqrt{\sqrt{\rho}\sigma\sqrt{\rho}})$~\cite{petz2011a,scandi2024}. In the following discussion, we primarily adopt this metric, with results for alternative choices provided in Supplemental Material~\cite{supp}.

We first focus on the case where the quantum state $\rho(t)$ evolves in a way such that it is differentiable with respect to time, as is the case for Markovian systems. Therefore, the trajectory length traced by the state over time is:
\begin{eqnarray}
\label{eq2}
\ell(t) &=& \frac{1}{2} \int_{0}^{t}ds = \frac{1}{2}\int_{0}^{t}\sqrt{\mathcal{D}(\rho, \rho+d\rho)} \\ \nonumber
 &=& \frac{1}{2} \int_{0}^{t}\sqrt{D\left(\rho, \partial_t\rho\right)}dt, \\ 
D(\rho,\partial_t\rho) & = & \text{Tr}\left[\partial_t\rho \mathcal{J}_{(x+1)/2}^{-1}|_{\rho}\left(\partial_t\rho\right)\right] \\ \nonumber
&=& \sum_{\{i,j|p_i + p_j \neq 0\}}\left|\langle i|\partial_t\rho|j\rangle\right|^2\frac{2}{p_i+p_j},
\end{eqnarray} 
with $\partial_t\rho\equiv\partial\rho/\partial t$.
We assume that $ \rho(t) = \sum_{i}p_i\left | i  \right \rangle \left \langle i \right | $ and omit the explicit time dependence for brevity. For evolution that leads to a unique steady state $\rho_{\text{steady}}$, we define the total trajectory length as $L\equiv\ell(t \to \infty)$. Therefore, with knowledge of the dynamics, $L$ can quantify the trajectory distance from a given initial state to $\rho_{\text{steady}}$. For convenience, we also define a residue distance $R(t) \equiv L -\ell(t)$. The IQME is characterized by considering two initial states, $\rho_A$ and $\rho_B$, where $R_{A}(0) > R_{B}(0)$. If there exists a finite time $t_c$ such that $R_{A}(t) < R_{B}(t)$ for all $t \ge t_c$, we identify this behavior as the IQME. In contrast, the traditional QME is based solely on the geodesic distance. Formally, we define the QME as a crossing point of $d(t) \equiv (1/2)\mathcal{D}^{\text{geo}}(\rho(t), \rho_{\text{steady}})$. If the dynamics impose every state $\rho$ to evolve precisely along the geodesic connecting it to $\rho_{\text{steady}}$, the QME and IQME are equivalent. However, in general, most states do not evolve along the geodesic, as the underlying dynamics are not typically fine-tuned to this path.

\begin{figure}[t]
\begin{center}
\includegraphics[width=3.4in, clip=true]{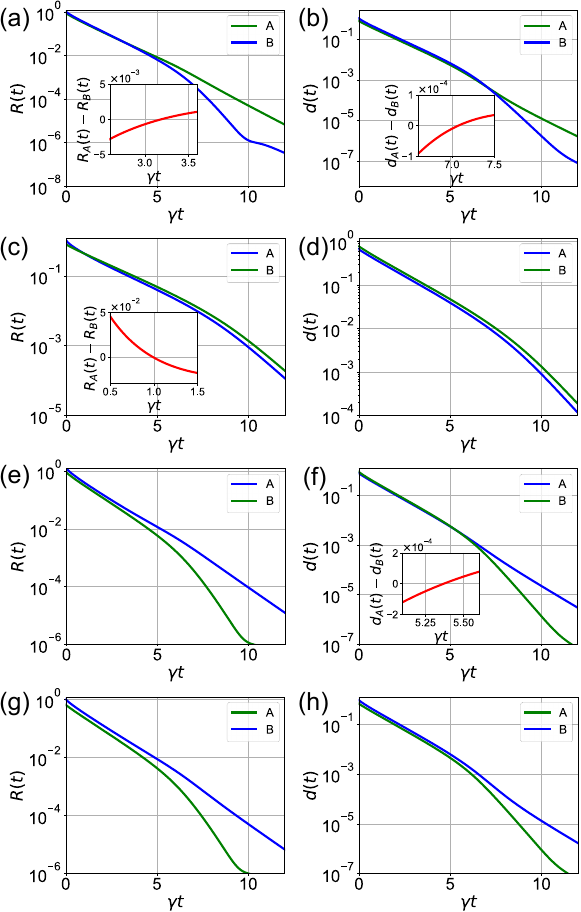}
\end{center}
\caption{IQME and QME in Markovian systems. The left column shows $R(t)$, where a crossing signifies the presence of IQME, while the right column shows $d(t)$, where a crossing indicates QME. Insets display $R_A(t)-R_B(t)$ and $d_A(t)-d_B(t)$ at the point of crossing, highlighted by the red curve crossing zero. We label the geodesically closer point as $A$ and depict the longer trajectory in blue in all cases. (a,b) Parameters correspond to case (i) in Fig.~\ref{fig1}. $(y_A(0), z_A(0))=(0.5, 0.0)$, $(y_B(0), z_B(0))=(0.0, 0.5)$. (c,d) Parameters correspond to case (ii). $(y_A(0), z_A(0))=(-0.95, -0.25)$, $(y_B(0), z_B(0))=(0.0, 0.0)$. (e,f) Parameters correspond to case (iii). $(y_A(0), z_A(0))=(0.9, 0.0)$, $(y_B(0), z_B(0))=(0.0, 0.2)$. (g,h) Parameters correspond to case (iv). $(y_A(0), z_A(0))=(0.0, -0.25)$,  $(y_B(0), z_B(0))=(0.5, 0.25)$.}
\label{fig2}
\end{figure}

\emph{Markovian system.---}To demonstrate the presence of IQME and better elucidate its counterintuitive origins, we consider a simple model in which an inverse QME has been both theoretically predicted and experimentally observed~\cite{shapira2024}. Specifically, we examine a single qubit whose dynamics are governed by the quantum Markovian master equation~\cite{manzano2020}: 
\begin{equation}
\label{eq4}
 \partial_t \rho = \mathcal{L}[\rho], 
\end{equation}
where $\mathcal{L}$ denotes the Lindbladian of the system. Following Ref.~\cite{shapira2024}, we have:
\begin{equation}
\frac{\partial \rho}{\partial (\gamma t)} =  -\frac{i}{2\gamma'}\left[\sigma_x, \rho\right]
+  \alpha \mathcal{A}_{\left | \uparrow  \right \rangle \left \langle \downarrow \right |}[\rho]
+ (1 - \alpha)\mathcal{A}_{\left | \uparrow  \right \rangle \left \langle \uparrow \right |}[\rho], 
\end{equation}
where $\mathcal{A}_{O}\left[\rho\right]\equiv O\rho O^{\dagger}-\frac{1}{2}\{O^{\dagger}O, \rho\}$. The dynamics are fully determined by $\gamma'$ and $\alpha$, which represent the relative strength between the total decoherence rate $\gamma$ and the unitary evolution, as well as the relative contributions of decay and dephasing, respectively. By parameterizing the state with the Bloch vector $\rho = (1/2)(1 + \vec{r} \cdot \vec{\sigma})$, where $\vec{r} = (x, y, z)$, we can easily visualize the evolution. It has been shown that the evolution in the $x$-direction is completely decoupled from the other two directions, with a stable fixed point at $x_{\text{steady}}=0$~\cite{shapira2024}. As a result, the state evolution can be clearly represented in the $y$-$z$ plane. As illustrated in Fig.~\ref{fig1}(b-e), we present four representative cases corresponding to those in Fig.~\ref{fig1}(a). For clarity, the state with a smaller geodesic distance to the steady state is labeled $A$, while the trajectory with a longer distance $L$ is shown in cyan in all cases.  For instance, in Fig.~\ref{fig1}(c), state $\rho_A$ has a shorter geodesic distance to $\rho_\text{steady}$, yet a longer trajectory distance, calculated using Eq.~(\ref{eq2}). The background heatmap illustrates the distribution of $L - d(0)$, highlighting the discrepancy between the actual trajectory and the geodesic path. This reveals that many states significantly deviate from the geodesic trajectory.

We further show the $R(t)$ and $d(t)$ for each case in Fig.~\ref{fig2}. Numerical details can be found in~\cite{supp}. It is evident that IQME manifests in the first two cases, while absent in the latter two. Moreover, IQME and QME can occur independently of each other, and cases (ii) and (iii) provide an insightful contrast. In case (ii), although the geodesically closer state $\rho_A$ follows a longer path, it reaches the steady state in a shorter time—a peculiarity captured by IQME but trivial under QME. In contrast, in case (iii), $\rho_A$ takes a longer time to reach the steady state, as intuition would suggest for a longer path. Here, IQME does not occur, but QME is observed. It is worth noting that previous studies attempting to explain the QME through information geometry have primarily focused on this specific case~\cite{bettmann2024,srivastav2024}. Thus, the IQME offers a more precise reflection of the counterintuitive nature of the dynamics.

As shown in Fig.~\ref{fig1} and Fig.~\ref{fig2}, the IQME bears a striking resemblance to the classical Brachistochrone problem. In the classical Brachistochrone scenario, dynamics are governed by Newton’s second law, $d^2\vec{r}/dt^2 = \vec{F}/m$, whereas in the quantum setting, the evolution is dictated by the master equation Eq.~(\ref{eq4}). In classical mechanics, motion is influenced by external force fields and constraints, while in quantum system, the Lindbladian $\mathcal{L}$ fully determines the dynamics. This analogy highlights how the IQME encapsulates counterintuitive behavior akin to that seen in the classical Brachistochrone problem. The misconception underlying the IQME arises from a similar oversight as in the classical Brachistochrone problem, although the path is longer, passing through certain points with very high velocity along the longer path can lead to a higher overall average speed, thus resulting in the behavior described by the IQME. Further analysis of this idea is provided in Supplemental Material~\cite{supp}.

\begin{figure}[t]
\begin{center}
\includegraphics[width=3.4in, clip=true]{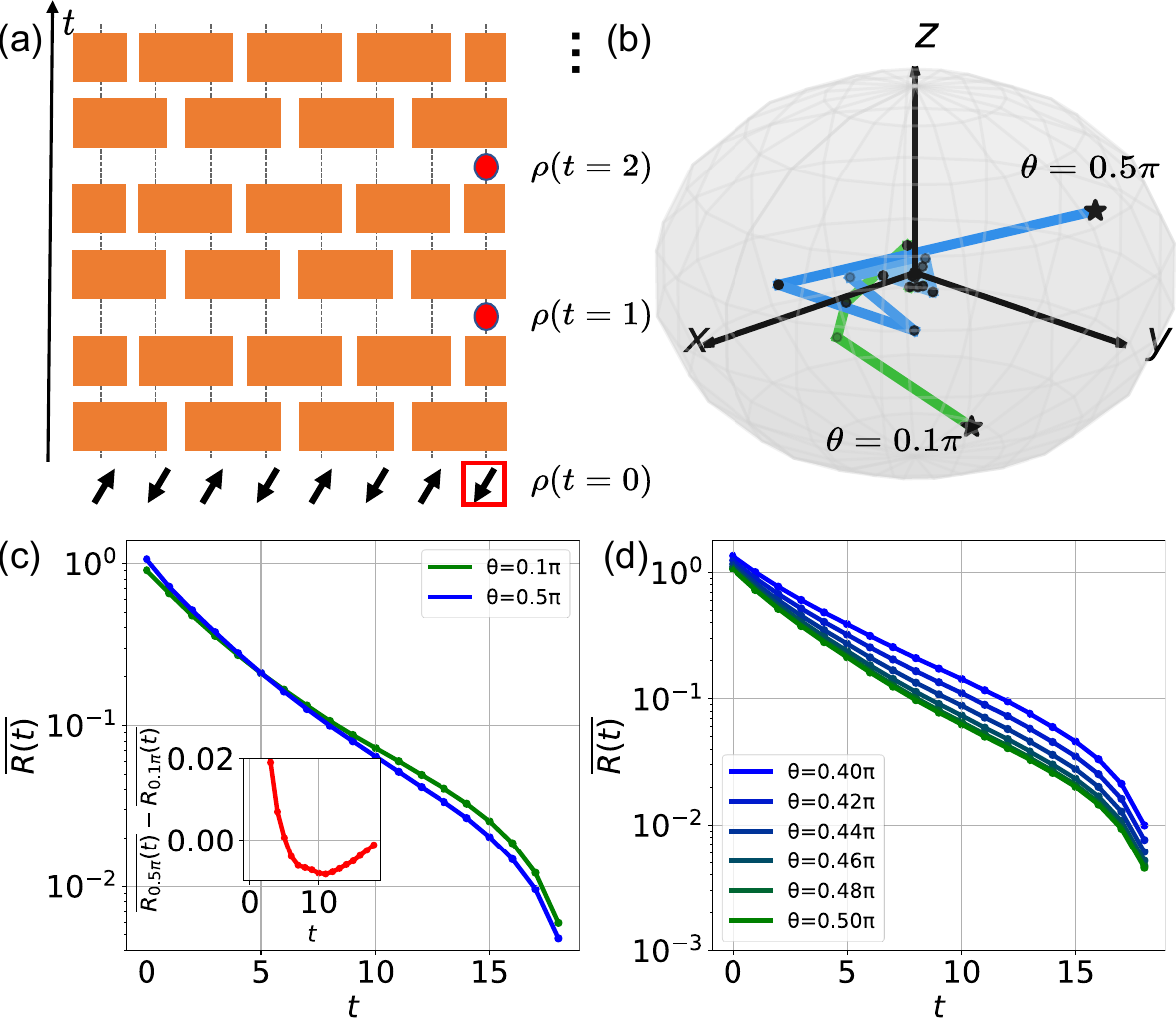}
\end{center}
\caption{IQME in a $U(1)$-symmetric quantum circuit. We consider a system of $N=16$ qubits and averaged over $10,000$ trajectories. (a) The quantum circuit comprises layers of $U(1)$-symmetric two-qubits gates arranged in a brick-wall pattern, as depicted by orange rectangles. We focus on the evolution of a single qubit as discussed in the main text. (b) State evolution for tilted N\'eel initial states with $\theta=0.1\pi$ and $\theta=0.5\pi$, where the color fades along the direction of the evolution. (c) $\overline{R(t)}$ for tilted N\'eel initial states with $\theta=0.1\pi$ and $\theta=0.5\pi$. (d) $\overline{R(t)}$ for tilted ferromagnetic initial states.}
\label{fig3}
\end{figure}

\emph{Quantum circuit.---}We now extend the definition of IQME to a quantum circuit, a closed system where a subsystem thermalizes due to the chaotic dynamics. Specifically, we consider a quantum circuit composed of random $U(1)$-symmetric two-qubit gates arranged in a brick-wall fashion, as illustrated in Fig.~\ref{fig3}(a). Each two-layer sequence corresponds to a single time step. The system's $U(1)$ charge operator is $\hat{Q} = \sum_i\sigma_z^{i}$, and each unitary $U$ in the circuit satisfies $[U, \hat{Q}] = 0$. The thermal equilibrium state is given by $\rho_{\text{eq}} \propto \text{exp}(-\lambda \hat{Q})$, where $\lambda$ depends on the expectation value of $\hat{Q}$ in the initial state. For a subsystem $M$, the state $\rho_M(t)$ converges to $\text{Tr}_{M_c}(\rho_{\text{eq}})$, where $M_c$ denotes the complement of subsystem $M$, as a result of thermalization~\cite{srednicki1994,dlessio2016,nandkishore2015,abanin2019,chang2024}. Since the evolution is no longer differentiable, we generalize $\ell(t)$ for a single trajectory as:
\begin{equation}
\ell(t_j) = \frac{1}{2}\sum_{i=0}^{j-1} \mathcal{D}^{\text{geo}}(\rho_{M}(t_{i+1}),\rho_{M}(t_{i})),
\end{equation}
which is a discretized version of the integral in Eq.~(\ref{eq2}). Due to the randomness in unitary gates, the evolution varies across different trajectories. Therefore, we average over different trajectories as $\overline{\ell(t_j)} = \mathbb{E}_{U}[\ell(t_j)]$. Importantly, this averaging is performed by first calculating the trajectory length for each individual path and then averaging these lengths, rather than taking the length of an averaged trajectory. This yields the averaged trajectory distance $\overline{L} = \overline{\ell(t_j \to \infty)}$ and the residue trajectory distance $\overline{R(t_j)} = \overline{L} - \overline{\ell(t_j)}$, with the crossing of $\overline{R(t_j)}$ defining IQME in this context. In the following, we choose a single qubit as the subsystem $M$, with additional numerical results available in~\cite{supp}.

We first examine tilted N\'eel initial states, parametrized as $ \left | \psi (\theta) \right \rangle = \text{exp}(-i\theta / 2\sum_i\sigma_{y}^{i})\left |\psi_0\right \rangle$, where $\left |\psi_0\right \rangle = \left |\uparrow \downarrow\uparrow \downarrow...\uparrow \downarrow\right \rangle $. As shown in Fig.~\ref{fig3}(b), $\rho_M$ approaches the maximally mixed state regardless of $\theta$, resembling the Markovian system with a single steady state. While the geodesic distance between the initial and final states is independent of $\theta$, we find that the trajectory distance increases as $\theta$ approaches $0.5\pi$. Notably, IQME is observed in this case, as demonstrated in Fig.~\ref{fig3}(c). Since the dynamics here are constrained solely by symmetry and locality, this phenomenon is expected to be universal, in contrast to the Markovian case, where dynamics depend on the specific Lindbladian. Next, we consider $\left |\psi_0\right \rangle$ as a ferromagnetic state. Strong finite-size effects prevent convergence to equilibrium for $\theta$ far from $0.5\pi$, leading to divergence in $\overline{L}$, so we focus on $\theta$ near $0.5\pi$~\cite{liu2024c}. In this case, the equilibrium state varies with $\theta$, taking the form $\rho_M(t\to \infty) \propto \text{exp}(-\lambda \sigma_z)$ where $\text{tanh}{\lambda} = -\text{cos}(\theta)$. No IQME is observed in ferromagnetic states, as shown in Fig.~\ref{fig3}(d). Interestingly, the QME, defined by entanglement asymmetry, manifests in an opposite manner~\cite{liu2024c,turkeshi2024}. For the tilted ferromagnetic state, the QME arises due to the shorter trajectory associated with more asymmetric initial states. Conversely, for tilted N\'eel states, more asymmetric initial states follow a longer trajectory. While symmetry restoration is slower for these states, they traverse the trajectory at a higher speed. This indicates that the rate of symmetry restoration does not necessarily correspond to the evolution speed along the trajectory.

\emph{Discussions.---}In contrast to QME, making analytical predictions for the occurrence of IQME is significantly more challenging, as deriving an analytical expression for the trajectory distance remains elusive. Nevertheless, we anticipate the IQME to emerge in general Markovian systems, as the model used in this work does not rely on specific or unique features. For the IQME observed in the $U(1)$-symmetric quantum circuit, the mechanisms underlying a longer yet faster traversal through Hilbert space for more asymmetric tilted N\'eel initial states warrant further investigation, which could potentially attributed to enhanced charge fluctuations~\cite{rylands2024}.

A natural extension of the IQME is to explore its applicability in non-Markovian systems, where QME has already been theoretically predicted~\cite{strachan2024}. If the dynamics in non-Markovian systems can be approximated to remain differentiable, the same IQME definition used for Markovian settings can be applied. Investigating whether the IQME manifests in such systems—and whether it can occur independently of the QME—represents a compelling direction for future research. Another promising avenue for exploring the IQME lies in quenched dynamics in closed systems. For instance, studies have shown that QME may arise in many-body localized systems, where dynamics are averaged over random Hamiltonians, akin to random quantum circuits~\cite{liu2024d}. By employing the IQME definition developed for quantum circuits, we can investigate its potential manifestation in these systems. These intriguing possibilities offer fertile ground for future investigations.

It is noteworthy that IQME can be experimentally verified with current technologies, provided full state tomography is achievable at each time step. An important and related direction for future exploration is whether the IQME can offer insights into the evolution rates of specific operator expectation values. In practical scenarios, interest often centers on a select subset of observable quantities rather than the entire quantum state, enabling experimental access without the significant overhead of full state tomography.

\begin{acknowledgments}
\emph{Acknowledgment.} This work is supported by the Natural Science Foundation of China through Grants No.~12350404 and No.~12174066, the Innovation Program for Quantum Science and Technology through Grant No.~2021ZD0302600, the Science and Technology Commission of Shanghai Municipality under Grants No.~23JC1400600 and No.~2019SHZDZX01.
\end{acknowledgments}

\end{document}


\title{Supplemental Material for ``Intrinsic Quantum Mpemba Effect in Markovian Systems and Quantum Circuits''}
\author{Dongheng Qian}
\affiliation{State Key Laboratory of Surface Physics and Department of Physics, Fudan University, Shanghai 200433, China}
\affiliation{Shanghai Research Center for Quantum Sciences, Shanghai 201315, China}
\author{Huan Wang}
\affiliation{State Key Laboratory of Surface Physics and Department of Physics, Fudan University, Shanghai 200433, China}
\affiliation{Shanghai Research Center for Quantum Sciences, Shanghai 201315, China}
\author{Jing Wang}
\thanks{wjingphys@fudan.edu.cn}
\affiliation{State Key Laboratory of Surface Physics and Department of Physics, Fudan University, Shanghai 200433, China}
\affiliation{Shanghai Research Center for Quantum Sciences, Shanghai 201315, China}
\affiliation{Institute for Nanoelectronic Devices and Quantum Computing, Fudan University, Shanghai 200433, China}
\affiliation{Hefei National Laboratory, Hefei 230088, China}


\maketitle

\tableofcontents

\begin{figure}[t]
\begin{center}
\includegraphics[width=5.8in, clip=true]{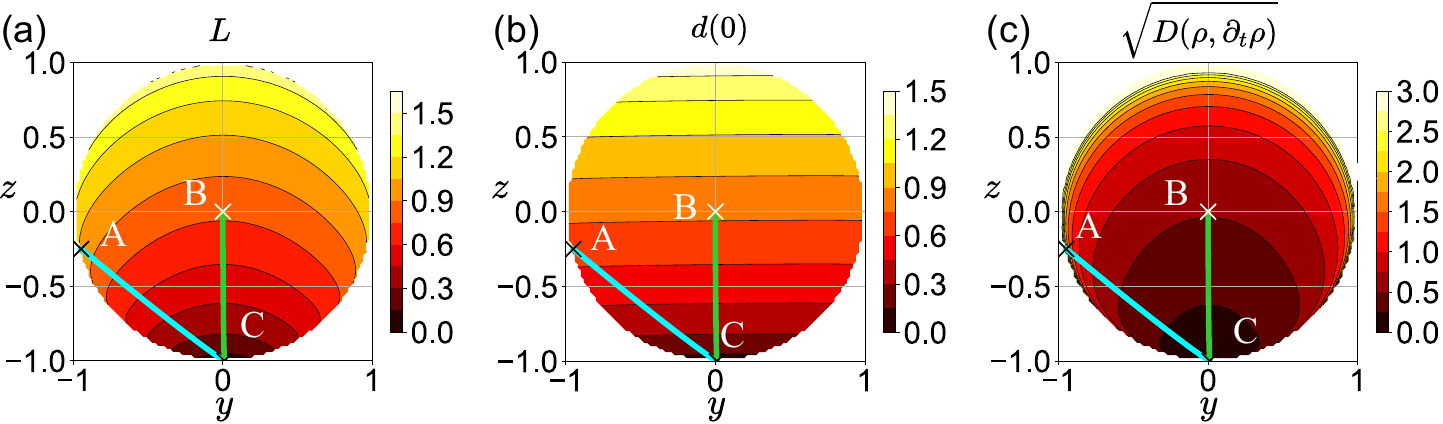}
\end{center}
\caption{Detailed examination of the IQME in Markovian systems. We take $\alpha=100$ and $\gamma' = 0.52$. Initial states $\rho_{A}$, $\rho_{B}$ corresponding to case (ii) in the main text. (a) Distribution of total trajectory lengths. (b) Distribution of geodesic distances. (c) Distribution of instantaneous speeds, with speeds greater than $3$ clipped for visual clarity.}
\label{fig1}
\end{figure}

\begin{figure}[t]
\begin{center}
\includegraphics[width=5.8in, clip=true]{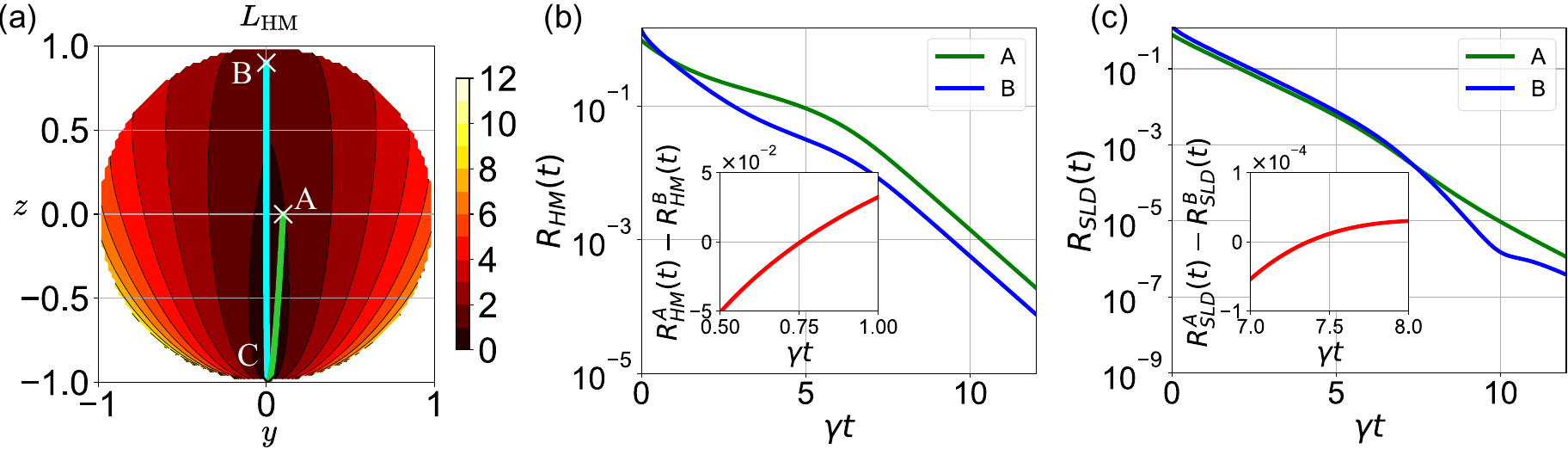}
\end{center}
\caption{IQME with HM metric. We choose initial states $\rho_A$ to be $(y_A(0), z_A(0)=(0.1, 0.0)$ and $\rho_B$ to be $(y_A(0), z_A(0)=(0.0, 0.9)$, with $\alpha=100$ and $\gamma' = 0.94$. (a) Distribution of $L_{\text{HM}}=\ell_{\text{HM}}(t \to \infty)$. (b) IQME is observed as a crossing in $R_{\text{HM}}(t)$. (c) For the same initial states $\rho_A$ and $\rho_B$, IQME is also observed under the SLD metric.}
\label{fig2}
\end{figure}

\section{Markovian systems}

\subsection{Numerical details}
In this section, we provide numerical details and additional results for investigating the IQME in Markovian systems. In the numerical simulations, the trajectory $\rho(t)$ is analytically tractable, as the Lindbladian is explicitly specified. To compute $\ell(t)$, we perform numerical integration using SciPy library in Python~\cite{virtanen2020} and the continuous quantum state trajectory is discretized into typically $1,000$ time steps. For all parameters considered in our work, $\ell(t)$ reaches convergence for $\gamma t \gtrsim 25$; therefore, we approximate the total trajectory length as $L = \ell(t=30)$.

To further illustrate that most states do not evolve along the geodesic, we present the distributions of $L$ and the geodesic distance $d(0)$ in Fig.~\ref{fig1}(a) and~\ref{fig1}(b), respectively, using $\alpha=100$ and $\gamma' = 0.52$ as an example. It is evident that these distances differ significantly for most states, except for those located near the $y=0$ axis. Additionally, we show the distribution of $\sqrt{D(\rho, \partial_t \rho)}$ in Fig.~\ref{fig1}(c), which represents the instantaneous speed of the state. While $\rho_A$ is initially farther from $\rho_C$ than $\rho_B$,  it more frequently traverses regions with higher instantaneous speeds compared to $\rho_B$, thereby explaining the counterintuitive IQME phenomenon. Analogous to the Brachistochrone problem, the fastest route emerges from a delicate balance between minimizing trajectory length and optimizing the ``steepness'' of the trajectory.

We focus on the Bures metric in the main text, also known as the symmetric logarithmic derivative (SLD) metric, with $f_{\text{SLD}}(x)=(x+1)/2$. Its corresponding geodesic distance is analytically related to the Uhlmann fidelity. It is natural to explore whether IQME persists under alternative metrics. It has been proven that all valid matrix monotone functions $f(x)$ are pointwise bounded as~\cite{scandi2024}:
\begin{equation}
\frac{2x}{x+1} \leq f(x) \leq \frac{x+1}{2}.
\end{equation}
The SLD metric corresponds to the largest monotone function, yielding the smallest distance metric between states, while the harmonic mean (HM) metric, defined as $f_{\text{HM}}(x)=2x/(x+1)$, represents the smallest monotone function and provides the largest distance metric. We first present the distribution of total trajectory length $L_{\text{HM}}$ using the HM metric in Fig.~\ref{fig2}(a) and confirm the occurrence of IQME under this metric in Fig.~\ref{fig2}(b). The two initial states also exhibit IQME when the SLD metric is used, as demonstrated in Fig.~\ref{fig2}(c). However, it is worth noticing that the occurrence of IQME for a given pair of initial states, $\rho_A$ and $\rho_B$, may depend on the chosen metric. Since the trajectory length $\ell(t)$ under any metric is bounded as $\ell_{\text{SLD}}(t) < \ell(t) < \ell_{\text{HM}}(t)$, it is attempting to define a universal IQME as a scenario where, although $R_{\text{SLD}}^{A}(0) > R_{\text{HM}}^{B}(0)$, there exists a finite time $t_c$ such that $R_{\text{HM}}^{A}(t) < R_{\text{SLD}}^{B}(t)$ for all $t > t_c$. Here, $R_{\text{X}}$ refers to the residue distance corresponding to the metric $\text{X}$. It is important to emphasize that the simultaneous occurrence of IQME under both the SLD and HM metrics does not necessarily imply the existence of a universal IQME. Conversely, the existence of a universal IQME guarantees the occurrence of IQME for any specific choice of metric. Although we find no evidence for universal IQME in this particular model, it remains an open question whether it might arise in other models and what conditions are necessary for its occurrence. 

\subsection{Relation to quantum speed limit}
It is also insightful to compare our findings with studies on the quantum speed limit (QSL), which addresses the maximal speed at which one state can evolve into another~\cite{delcampo2013, taddei2013, funo2019, sekiguchi2024,vanvu2023}. The general QSL imposed by the geometry utilizes the fact that the trajectory length connecting $\rho(0)$ and $\rho(\tau)$ must be at least as long as the geodesic distance, i.e., $\ell(\tau) \geq 2 \mathcal{D}^{\text{geo}}(\rho(0), \rho(\tau))$. This bound is saturated only when the trajectory precisely follows the geodesic. If an explicit expression for $\ell(\tau)$ is available, it can serve to bound the evolution time, with the minimum time typically achieved when the trajectory aligns with the geodesic. Consequently, it is argued that the optimal CPTP map transporting state $\rho(0)$ to $\rho(\tau)$ follows the geodesic path. However, the IQME differs in that it compares travel times for different initial states reaching the same final state, rather than comparing time costs for a single initial state under varying dynamics. Thus, even if $\rho_B$ travels along the geodesic while $\rho_A$ significantly deviates from it, $\rho_A$ may still reach the steady state faster, as demonstrated in case (ii) in the main text.

\section{Quantum circuits}
\begin{figure}[t]
\begin{center}
\includegraphics[width=5.8in, clip=true]{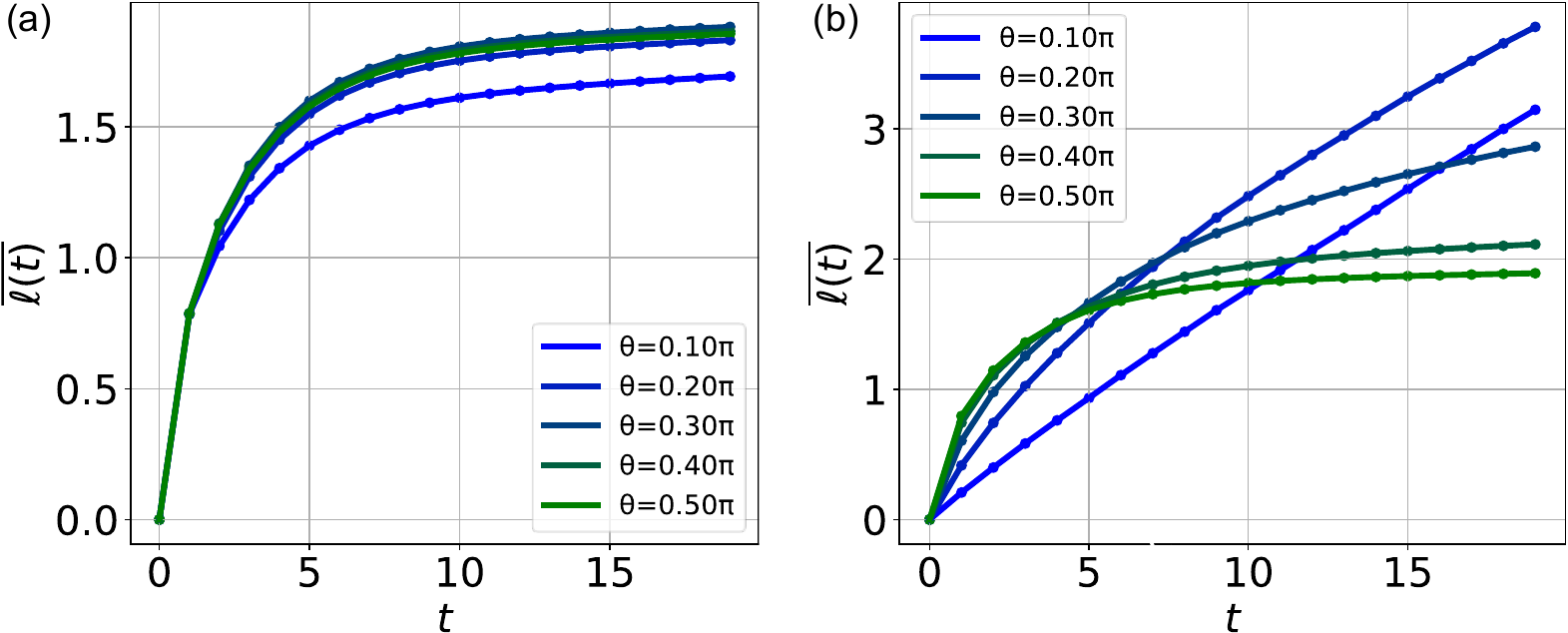}
\end{center}
\caption{Trajectory distances $\overline{\ell(t)}$ for different initial states. Every point is averaged over $10,000$ trajectories. (a) Tilted N\'eel initial states. (b) Tilted ferromagnetic initial states. }
\label{fig3}
\end{figure}

\begin{figure}[b]
\begin{center}
\includegraphics[width=5.8in, clip=true]{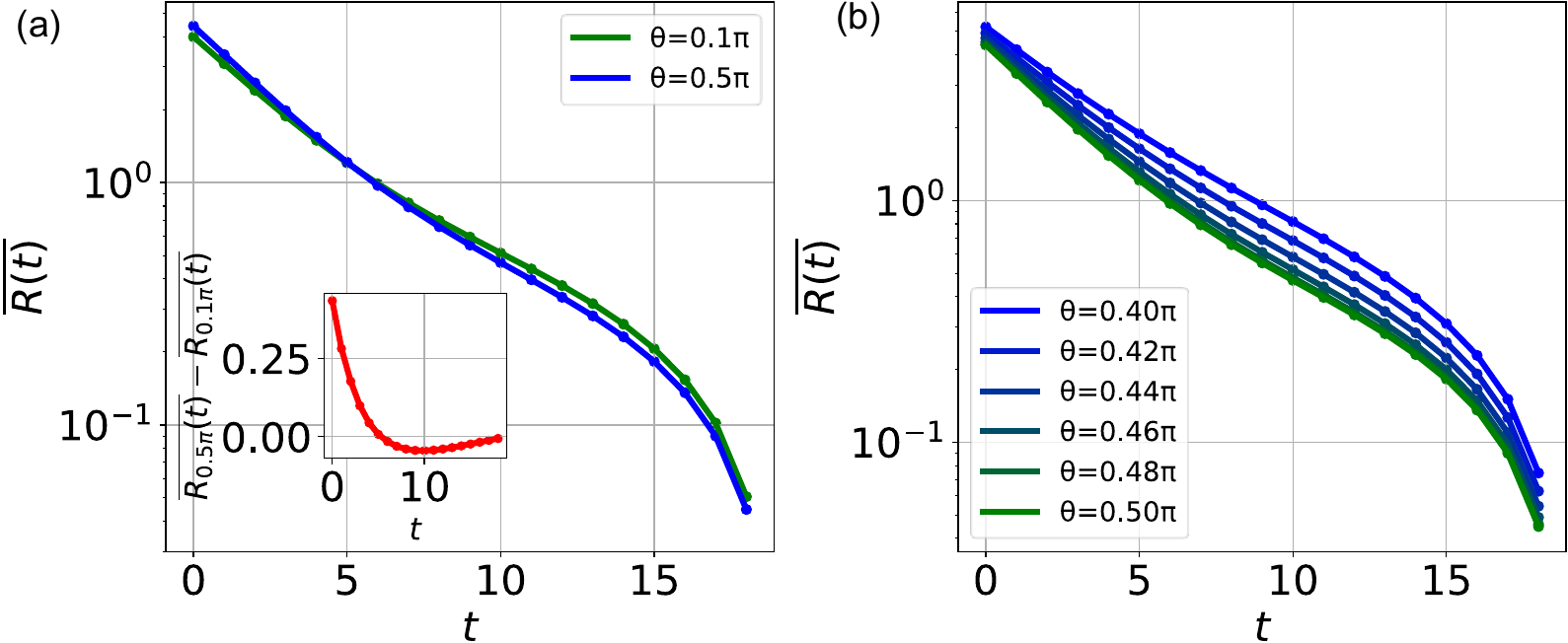}
\end{center}
\caption{IQME for subsystem consists of $N/4=4$ qubits, with a total system size of $N=16$. Every point is averaged over $10,000$ trajectories. (a) Tilted N\'eel initial states, with the crossing highlighted in the inset for clarity. (b) Tilted ferromagnetic initial states. }
\label{fig4}
\end{figure}

\begin{figure}[t]
\begin{center}
\includegraphics[width=5.8in, clip=true]{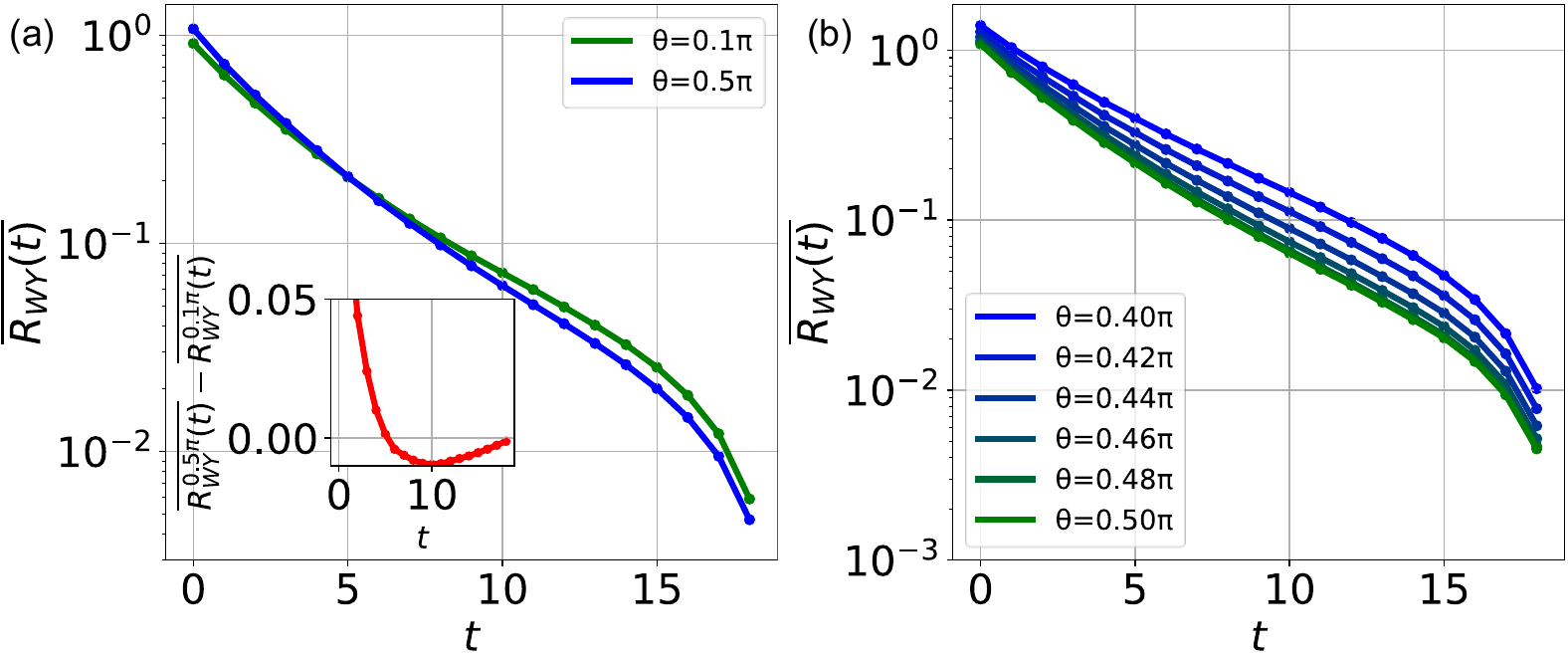}
\end{center}
\caption{IQME for WY metric. Every point is averaged over $10,000$ trajectories. (a) Tilted N\'eel initial states, with the crossing highlighted in the inset for clarity. (b) Tilted ferromagnetic initial states. }
\label{fig5}
\end{figure}

\begin{figure}[b]
\begin{center}
\includegraphics[width=5.8in, clip=true]{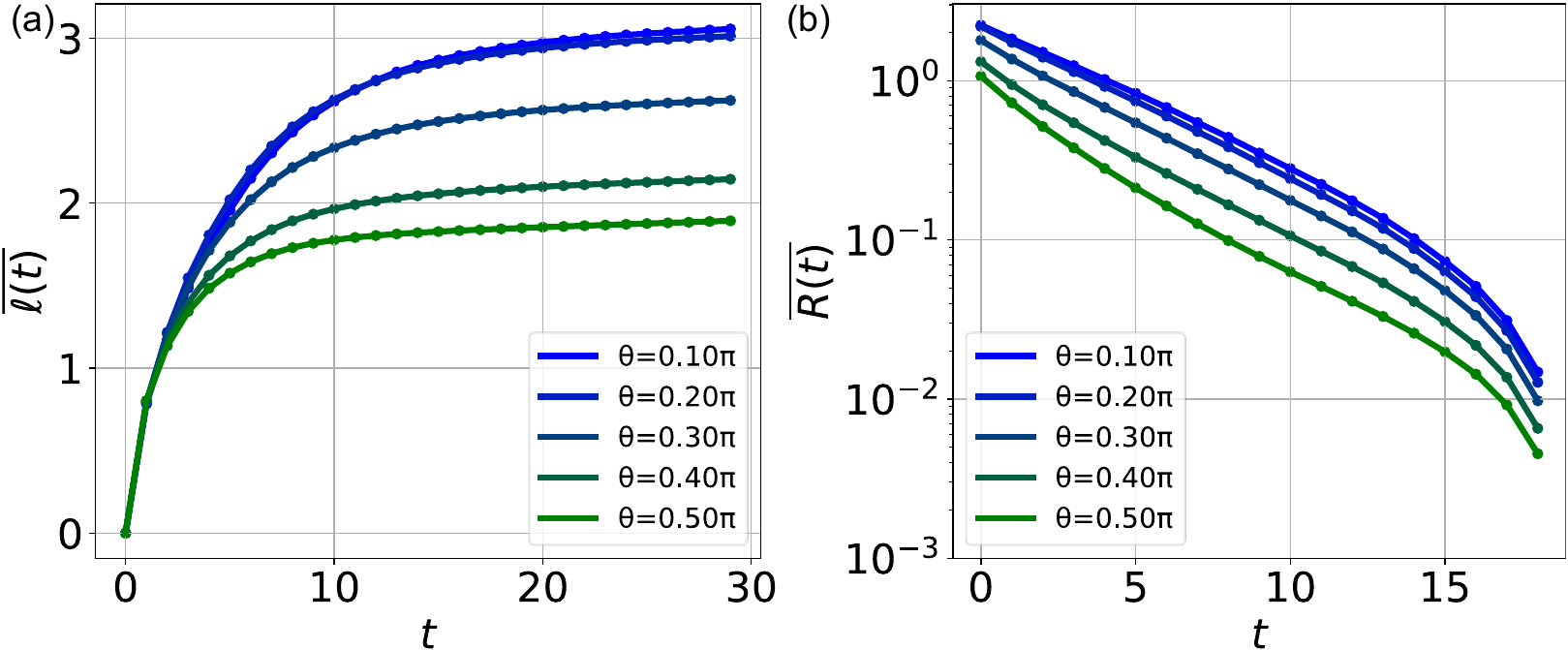}
\end{center}
\caption{IQME for tilted ferromagnetic initial states with a domain wall in the middle. Every point is averaged over $10,000$ trajectories. (a) The trajectory distance converges in this case, in contrast to the behavior observed for initial states without a domain wall. (b) IQME is not present, as evidenced by the absence of a crossing in $\overline{R(t)}$.}
\label{fig6}
\end{figure}

In this section, we present numerical details and additional results for the investigation of IQME in $U(1)$-symmetric quantum circuits. Each $U(1)$-symmetric two-qubit gate can be block-diagonalized into three blocks: $U_{1\times1}^{(1)}$, $U_{2\times2}^{(0)}$ and $U_{1\times1}^{(-1)}$, corresponding to total charge $1$, $0$ and $-1$, respectively. Each block matrix is randomly sampled from the Haar measure. All numerical simulations are performed under periodic boundary conditions, making the system translational invariant. To compute $\overline{L}$, we first determine the convergence time for different initial states. As shown in Fig.~\ref{fig3}(a), the evolution converges around $t=15$ for tilted N\'eel initial states. The small slope observed after convergence is attributed to fluctuations around the steady state induced by the randomness of the unitary gates. Therefore, we approximate $\overline{L}$ to be $\overline{\ell(t=20)}$. 

For tilted ferromagnetic initial states, the evolution of $\overline{\ell(t)}$ is shown in Fig.~\ref{fig3}(b). For small tilted angles $\theta$, the reduced density matrix for a single qubit does not reach the thermalized steady state, as indicated by a significant slope in $\overline{\ell(t)}$. This observation is consistent with the results in Refs.~\cite{liu2024c, turkeshi2024}, which analytically demonstrated the existence of a crossover between symmetry-restored and symmetry-breaking phases at small $\theta$ in finite-size systems. In contrast, such a crossover is absent for tilted N\'eel initial states. To mitigate finite-size effects, we focus on examining IQME for large tilt angles $\theta$. From a generalized perspective, we can regard $\overline{L} \to \infty$ for $\theta \lesssim 0.4\pi$ in a finite-size system. Consequently, $\overline{R(t)}$ for $\theta \lesssim 0.4\pi$ remains larger than that for $\theta \ge 0.4 \pi$, resulting in the absence of IQME. Combined with the results in the main text, this finding suggests that the observed QME in tilted ferromagnetic states arises from the shorter trajectory distances traversed by more asymmetric initial states.

In the main text, we focus on cases where the subsystem consists of a single qubit. Here, we extend to the situation where the subsystem comprises $N/4=4$ qubits, with the results presented in Fig.~\ref{fig4}. The IQME persists for tilted N\'eel initial states but is absent for tilted ferromagnetic initial states. Additionally, it is worth exploring whether IQME persists under other metrics. Besides the Bures metric, another metric with an analytically derived geodesic distance is the Wigner-Yanase (WY) metric, defined by $f_{\text{WY}}(x)=(1+\sqrt{x})^2/4$~\cite{gibilisco2003}. The geodesic distance for the WY metric is given by $\mathcal{D}^{\text{geo}}_{\text{WY}}=2 \text{arccos}\mathcal{A}$, where $\mathcal{A} = \text{Tr}(\sqrt{\rho}\sqrt{\sigma})$ is the quantum affinity~\cite{scandi2024, bettmann2024}. Results for the WY metric are shown in Fig.~\ref{fig5}, and the behavior remains the same with that for the Bures metric. Finally, we investigate tilted ferromagnetic initial states with a domain wall in the middle, where $\left | \psi_0 \right \rangle = \left | \uparrow \uparrow \uparrow ... \downarrow \downarrow \downarrow \right \rangle$. For these states, the reduced density matrix $\rho_M$ would as well approach the maximally mixed state irrespective of $\theta$. The QME, defined via entanglement asymmetry, has been observed in this scenario~\cite{liu2024c}. To analyze IQME, we first verify the convergence of the trajectory distance, as shown in Fig.~\ref{fig6}(a). Subsequently, we plot $\overline{R(t)}$ for different values of $\theta$ and find no evidence of IQME, as illustrated in Fig.~\ref{fig6}(b). Thus, the observed QME in this case can also be attributed to the shorter trajectory distances traversed by more asymmetric initial states.

\newpage